\documentclass[11pt,a4paper]{article}
\usepackage{moriond,epsfig}
\def\be{\begin{equation}}
\def\ee{\end{equation}}
\def\bea{\begin{eqnarray}}
\def\eea{\end{eqnarray}}

\def \bea{\begin{eqnarray}}
\def \beq{\begin{equation}}

\def \eea{\end{eqnarray}}
\def \eeq{\end{equation}}

\def \ob{\overline{B}^0}
\def \od{\overline{D}^0}

\def \tu{\tau({\rm Universe})}

\begin{document}
\vspace*{4cm}
\title{MORIOND QCD 2007 -- THEORY SUMMARY}

\author{JONATHAN L. ROSNER}
\address{Enrico Fermi Institute, University of Chicago \\
5640 South Ellis Avenue, Chicago, IL 60637, USA}
\maketitle\abstracts{Developments reported at the 2007 Moriond Workshop on QCD
and Hadronic Interactions are reviewed and placed in a theoretical context.}

\section{Introduction}
QCD was invented in 1973.  (There were some earlier hints.)  We are still
concerned with it as neither perturbative nor currently available
non-perturbative (e.g., lattice) methods apply to many interesting phenomena.
These include hadron structure, spectroscopy, jet and quarkonium
fragmentation, heavy ion physics, and effects of thresholds.  The understanding
of hadronic behavior is crucial in separating underlying short-distance physics
(whether electroweak or new) from strong-interaction effects.  The properties
of hadrons containing heavy quarks provide an exceptional window into QCD
tests.  Finally, QCD may not be the only instance of important non-perturbative
effects; familiarity with it may help us to prepare for surprises at the
Large Hadron Collider (LHC).  In this review we shall discuss a number of
developments reported at Moriond QCD 2007 in the context of these ideas.  A
companion review \cite{Rolandi} deals directly with the experimental results.
I apologize for not covering some theoretical topics whose relation to
experimental results presented at this conference is not yet clear to me,
and for omitting some nice experimental results for which I have no comments.

\section{Heavy flavor issues:  the current CKM matrix}
\begin{figure}
\begin{center}
\includegraphics[width=0.7\textwidth]{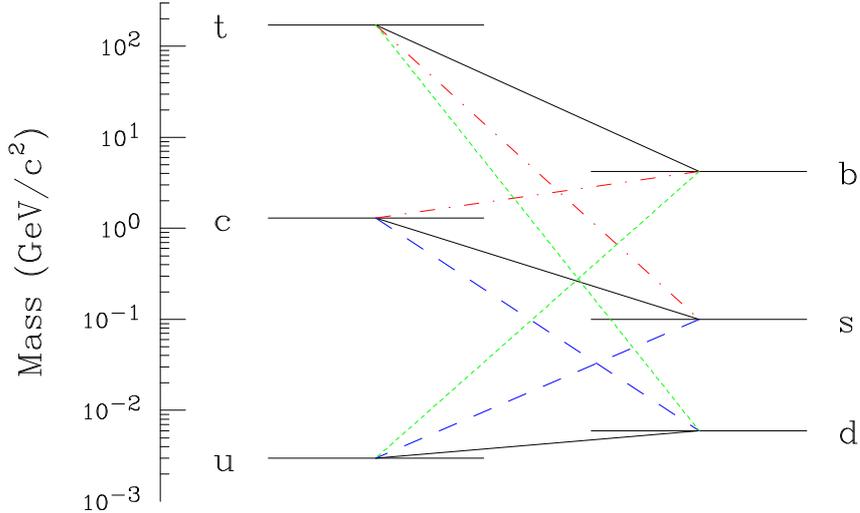}
\end{center}
\caption{The quarks and weak charge-changing transitions among them.  Solid,
dashed, dash-dotted, and dotted lines correspond to successively weaker
transitions.
\label{fig:quarks}}
\end{figure}

The Kobayashi-Maskawa matrix theory of CP violation, and its parametrization of
charge-changing weak transitions, as shown in Fig.\ \ref{fig:quarks}, passes
all experimental tests so far.  The major uncertainties in the parameters of
the CKM matrix are now dominated by theory.  Briefly, we have
$V_{ud} \simeq V_{cs} \simeq 0.974$, $V_{us} \simeq -V_{cd} \simeq 0.226$,
$V_{cb} \simeq -V_{ts} \simeq 0.041$, $V_{td} \simeq 0.008 e^{-i~21^\circ}$,
$V_{ub} \simeq 0.004e^{-i~66^\circ}$ (sources of phase information will be
explained below), and -- on the basis of single-top production observed by the
D0 collaboration \cite{D0top} -- $0.68 < |V_{tb}| < 1$ at 95\% c.l.  

\section{Meson decay constants and implications}

The ability of theory to anticipate important hadronic properties is
illustrated by recent results on meson decay constants.  Moreover, it has been
possible in some cases to replace calculated quantities with better-determined
experimental ones, reducing errors on fundamental parameters such as CKM
matrix elements.
 
In 2005 the CLEO Collaboration \cite{CLEOfD} reported the measurement $f_{D^+}
= (222.6 \pm 16.7^{+2.8}_{-3.4})$ MeV, to be compared with one lattice QCD
prediction \cite{lattfD} of $201\pm3\pm17$ MeV.  More recently CLEO has
measured $f_{D_s} = (274 \pm 13 \pm 7)$ MeV.~\cite{Artuso:2007zg}  [One can
obtain a slightly more precise value by including preliminary data on $D_s \to
\tau \nu$ where $\tau \to e \nu \bar \nu$.~\cite{Stone:ICHEP}]  The BaBar
Collaboration reports $f_{D_s} = (283 \pm 17 \pm 7 \pm 14)$ MeV.%
~\cite{Aubert:2006sd}

One lattice prediction \cite{lattfD} is $f_{D_s} = 249 \pm 3 \pm 16$ MeV,
leading to a predicted ratio $f_{D_s}/f_D = 1.24 \pm 0.01 \pm 0.07$.  This is
to be compared with the CLEO ratio $1.23 \pm 0.11 \pm 0.04$.%
~\cite{Artuso:2007zg}
One expects $f_{B_s}/f_B \simeq f_{D_s}/f_D$ so better measurements of
$f_{D_s}$ and $f_D$ by CLEO will help validate lattice calculations and
provide input for interpreting $B_s$ mixing.  A desirable error on $f_{B_s}/f_B
\simeq f_{D_s}/f_D$ is $\le 5\%$ for a useful determination of the CKM element
ratio $|V_{td}/V_{ts}|$.  This will require errors $\le 10$ MeV on $f_{D_s}$
and $f_D$.  (Independent information on $|V_{td}/V_{ts}|$ has come from a
precise measurement of $B_s$--$\overline{B}_s$ mixing.~\cite{Abulencia:2006ze})
A scaling argument from the quark model \cite{Rosner:1990xx}
implies $f_{D_s}/f_D \simeq f_{B_s}/f_B \simeq \sqrt{M_s/M_d} \simeq 1.25$,
with constituent masses $M_s \simeq 485$ MeV, $M_d \simeq 310$ MeV.

\section{$B_s$ physics}

Comparing box diagrams for $b \bar s \to s \bar b$ and $b \bar d \to d \bar b$
(dominated by intermediate top quarks), one sees that $B_s$--$\overline{B}_s$
mixing is stronger than $B$--$\overline{B}$ mixing because $|V_{ts}/V_{td}|
\simeq 5$.  Now, CKM unitarity implies $|V_{ts}| \simeq |V_{cb}| \simeq 0.041$
is well measured, so $B_s$--$\overline{B}_s$ mixing really probes the matrix
element between $B_s$ and $\overline{B}_s$.  This quantity involves $f_{B_s}^2
B_{B_s}$, whose ratio with respect to that for non-strange $B$'s is known from
lattice QCD:~\cite{Okamotoxi} $\xi \equiv f_{B_s} \sqrt{B_{B_s}}/(f_B
\sqrt{B_B}) = 1.21^{+0.047}_{-0.035}$.  The $B^0$--$\ob$ mixing amplitude is
well-measured:  $\Delta m_d = (0.507 \pm 0.004)$ ps$^{-1}$.  Consequently,
measurement of $B_s$ mixing implies a value of $|V_{td}/V_{ts}|$.  The recent
CDF measurement at Fermilab $\Delta m_s=(17.77 \pm 0.10 \pm 0.07)$ ps$^{-1}$
\cite{Abulencia:2006ze} gives $|V_{td}/V_{ts}| = 0.206 \pm 0.008$ and hence
$1 - \rho - i \eta \equiv |V^*_{tb}V_{td}/(V^*_{cb}V_{cd})| = 0.91 \pm 0.04$.
This implies that $\gamma \equiv {\rm Arg}(V^*_{ub}V_{ud}/(V^*_{cb}V_{cd})
\simeq (66 \pm 6)^{\circ}$, a great improvement over previous determinations.

The first evidence for $B_s$ mixing was presented by the D0 collaboration.%
~\cite{D0mix}  This collaboration has now presented evidence for a decay rate
difference between the $B_s$ mass eigenstates, with the eigenstate which is
approximately CP-even decaying somewhat more rapidly:~\cite{D0life}  $\Delta
\Gamma_s=0.13\pm0.09~{\rm ps}^{-1}$.  This agrees with the expected value%
~\cite{Lenz:2006hd}
$\Delta \Gamma_s \simeq (1/200) \Delta m_s \simeq 0.09$ ps$^{-1}$. (The
values of $\Delta \Gamma_s$ and $\Delta m_s$ are expected to track one
another.)  Within large errors, D0 sees no evidence for CP violation in $B_s
\to J/\psi \phi$.  One expects in the Standard Model $\phi_s = 0.036 \pm
0.003$, a value which may be accessible to LHCb.~\cite{Goloutvin,John}

\section{Systematics of $B$ decays}

\subsection{General considerations}

\begin{figure}
\includegraphics[height=2in]{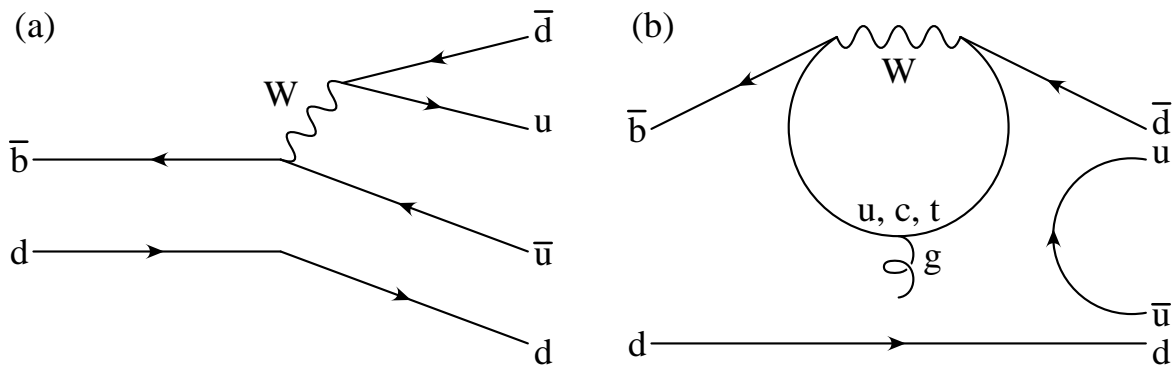}
\caption{Examples of decay topologies for $B^0 \to \pi^+ \pi^-$.  (a) Tree;
(b) penguin.
\label{fig:treepen}}
\end{figure}

Reviews of $B$ decays were given at this Conference by Lin \cite{Lin}
(experiment) and L\"u \cite{Lu} (theory).  It is useful to visualize $B$ decay
amplitudes in terms of flavor diagrams \cite{Gronau:1994rj} (see, e.g., Fig.\
\ref{fig:treepen}).  Flavor SU(3) permits one to relate decay asymmetries in
one channel to those in another.  For example, one can show \cite{DH,GR}
\beq
\Gamma(\bar B^0 \to \pi^+ \pi^-) - \Gamma(B^0 \to \pi^+ \pi^-) = -
[\Gamma(\bar B^0 \to K^- \pi^+) - \Gamma(B^0 \to K^+ \pi^-)]~.
\eeq
Using dominance of $B \to K \pi$ transitions by the isospin-preserving
($\Delta I = 0$) penguin $\bar b \to \bar s$ transition, and a well-established
hierarchy of other amplitudes, one can obtain sum rules for rates \cite{rates}
and asymmetries \cite{fourCPsr,threeCPsr} in these decays.  Defining the
CP-averaged ratios
\beq
R \equiv \frac{\bar\Gamma(B^0 \to K^+ \pi^-)}
               {\bar\Gamma(B^+ \to K^0 \pi^+)}~,~~
R_c\equiv \frac{2\bar\Gamma(B^+ \to K^+ \pi^0)}
               {\bar\Gamma(B^+ \to K^0 \pi^+)}~,~~
R_n\equiv \frac{\bar\Gamma(B^0 \to K^+ \pi^-)}
               {2\bar\Gamma(B^0 \to K^0 \pi^0)}
\eeq
where $\bar \Gamma(B\to f)\equiv [\Gamma(B\to f)+\Gamma(\bar B\to \bar f)]/2$,
one such sum rule is $R_c = R_n$.  Experimentally~\cite{hfag}
\beq
R=0.90 \pm 0.05~,~~R_c=1.11 \pm 0.07~,~~R_n=0.97 \pm 0.07~,
\eeq
so the sum rule is satisfied.  It is expected to hold also to first order
in isospin breaking.~\cite{ibreak}

A recent result is relevant to the systematics of $B \to PV$ decays, where $P$
and $V$ are light pseudoscalar and vector mesons.  The pure penguin process
$B^+ \to K^0 \rho^+$ has been seen by BaBar \cite{Aubert:2007mb} with a
branching
ratio ${\cal B}(B^+ \to K^0 \rho^+) = (8.0^{+1.4}_{-1.3} \pm 0.5) \times
10^{-6}$.  This is comparable to the pure-penguin process $B^+ \to K^{*0}\pi^+$
with ${\cal B} = (10.7 \pm 0.8) \times 10^{-6}$.  In the former process, the
spectator quark ends up in a vector meson (``$p_V$''), while in the latter the
spectator ends up in a pseudoscalar (``$p_P$'').  This confirms an early
expectation by Lipkin \cite{pVpP} that the amplitudes for the two processes
were related by $p_V \simeq - p_P$.

\subsection{$B_s$ decays}

One way to learn the width difference $\Delta \Gamma$ of $B_s$ mass
eigenstates is to compare the decay lifetimes in different polarization states
of the final vector mesons in $B_s \to J/\psi \phi$.  These are conveniently
expressed in a Cartesian basis.~\cite{Dighe:1995pd} There are three such
states.  Two are CP-even.  In one of these, the vector mesons' linear
polarizations are perpendicular to the decay axis and parallel to one another
(``$\parallel$'').  In the other CP-even state, both vector mesons are
longitudinally polarized (``0'').  In the CP-odd state, the vector mesons'
linear polarizations are perpendicular to the decay axis and also to one
another (``$\perp$'').  Separating out the CP-even and CP-odd lifetimes would 
be much easier using $\parallel$ and $\perp$ states, thereby avoiding
bias due to imperfect modeling of polar angle dependence.

The branching ratio ${\cal B}(B_s \to K^+ K^-) = (24.4 \pm 1.4 \pm 4.6) \times
10^{-6}$ reported by CDF at this Conference \cite{Scuri} is due mainly to the
$|\Delta S| = 1$ penguin.  For comparison, ${\cal B}(B^+ \to K^0 \pi^+) = (23.1
\pm 1.0) \times 10^{-6}$.  The large error on the former means that one can't
see the effects of non-penguin amplitudes through interference with the
dominant penguin.

$B_s$ decays help validate flavor-SU(3) techniques used in extracting CKM
phases.  For example, under the U-spin transformation $d \leftrightarrow s$,
the decay $B_s \to K^- \pi^+$ is related to $B^0 \to K^+ \pi^-$.  It has a
branching ratio of $(5.0 \pm 0.75 \pm 1.0) \times 10^{-6}$; it differs from
the process $B^0 \to \pi^+ \pi^-$ with ${\cal B} = (5.16 \pm 0.22) \times
10^{-6}$ only by having a different spectator quark.

\subsection{Baryonic $B$ decays}

Results presented at this conference \cite{Medvedeva:2007af,Wang:2007as,MZWang}
shed light on the
mechanisms of $B$ decays to baryonic final states.  Low-mass baryon-antibaryon
enhancements seen in these decays favor a fragmentation picture over resonant
substructure, based in part on information from angular correlations between
decay products.  The production of several heavy quarks, as in $b \to c s \bar
c$, helps produce baryons like $csq$ where $q=(u,d)$ gives $\Xi_c$ and $q=s$
gives $\Omega_c$.  The large available phase space and high quark multiplicity
in $B$ decays may permit the production of exotic final states.~\cite{JRexot}

\subsection{Sum rules for CP asymmetries in $B \to K \pi$}

Using the dominance of the $\Delta I = 0$ $\bar b \to \bar s$ penguin
amplitude, M. Gronau \cite{fourCPsr} has shown that
\beq
A_{CP}(K^+ \pi^-) + A_{CP}(K^0 \pi^+) =
A_{CP}(K^+ \pi^0) + A_{CP}(K^0 \pi^0)~~.
\eeq
Non-penguin amplitudes should be small in $B^+ \to K^0 \pi^+$,
so $A_{CP}(K^0 \pi^+) \simeq 0$ and \cite{threeCPsr}
\beq
A_{CP}(K^+ \pi^-) = A_{CP}(K^+ \pi^0) + A_{CP}(K^0 \pi^0)~~.
\eeq
[Strictly speaking, a more accurate version of these sum rules applies to
CP-violating rate differences $\Delta(f) \equiv \Gamma(\bar B \to \bar f)
- \Gamma(B \to f)$.]
The observed CP asymmetries \cite{hfag} are $A_{CP}(K^+ \pi^-) = -0.097 \pm
0.012$, $A_{CP}(K^0 \pi^+) = 0.009 \pm 0.025$, $A_{CP}(K^+ \pi^0) = 0.047
\pm 0.026$, and $A_{CP}(K^0 \pi^0) = -0.12 \pm 0.11$.  The last is the most
poorly known and may instead be predicted using the sum rules. With corrections
for $\tau(B^+)/\tau(B^0) = 1.076 \pm 0.008$ and branching ratios, the first and
second of these sum rules predict $A_{CP}(K^0 \pi^0) = (-0.140 \pm 0.043,
-0.150 \pm 0.035)$.  The experimental value of $A_{CP}(K^0 \pi^0)$ carries too
large an error at present to provide a test.

A vanishing $A_{CP}(K^0 \pi^0)$ would imply $A_{CP}(K^+ \pi^-) =
A_{CP}(K^+ \pi^0)$, which is not so. $A_{CP}(K^+ \pi^0)$ and $A_{CP}(K^0
\pi^0)$ involve color-suppressed tree and electroweak penguin (EW) amplitudes.
The latter occur in a calculable ratio $\delta_{\rm EW} = 0.60 \pm 0.05$ with
respect to known amplitudes.

One may ask how the CP asymmetry in $B^0 \to K^+ \pi^-$ can be non-zero,
thereby signaling the presence of non-penguin amplitudes, while neither the
CP asymmetry nor the rate ratio $R_c$ shows evidence of such amplitudes in
$B^+ \to K^+ \pi^0$.  Let $r_c \sim 0.2$ denote the ratio of tree to penguin
amplitudes in $B^+ \to K^+ \pi^0$.  One may write the sum rule \cite{GRsr}
\beq
\left( \frac{R_c-1}{\cos \gamma - \delta_{\rm EW}} \right)^2 +
\left( \frac{A_{CP}(B^+ \to K^+ \pi^0)}{\sin \gamma} \right)^2 = (2r_c)^2 +
{\cal O}(r_c^3)~,
\eeq
which is essentially based on the identity $\cos^2 \delta + \sin^2 \delta = 1$,
where $\delta$ is a strong phase.  The key to this sum rule's validity is that
$\cos \gamma \simeq \delta_{\rm EW}$, thereby allowing it to be satisfied for
$R_c \simeq 1$ and small $A_{CP}(K^+ \pi^0)$.

\subsection{Ways to measure $\sin 2 \beta$}

The BaBar Collaboration has updated its value based on $b \to c \bar c s$
decays:~\cite{BaBsinbeta} $\sin 2 \beta = 0.714 \pm 0.032 \pm 0.018$.  When
combined with the latest Belle value \cite{Belsinbeta} of $0.642 \pm 0.031
\pm 0.017$ and earlier data this gives a world average \cite{hfag,Stocchi}
$\sin 2 \beta = 0.678 \pm 0.025$, serving as a reference for all other
determinations of $\beta$.

Recently BaBar studied in the decay $B^0 \to D_{CP}^{(*)0} h^0$, extracting
coefficients $S$ and $C$ of time-dependent decay rate modulations proportional
to $\sin \Delta m t$ and $\cos \Delta m t$.~\cite{BaDh}  The result
$\sin 2 \beta_{\rm eff} = - S = 0.56 \pm 0.23 \pm 0.05$ is compatible with
the reference value.  The value $C = -0.23 \pm 0.16 \pm 0.04$ is compatible
with no direct CP violation, as expected in the Standard Model, but carries a
large experimental error.

A large number of processes are dominated by $b \to s$ penguin amplitudes.
When averaged,~\cite{hfag} these give $\sin 2 \beta_{\rm eff} = 0.53 \pm 0.05$,
a value $2.6 \sigma$ below the reference value.  It is not clear that it makes
sense to average all these processes as some involve $b \to s \bar s s$, others
$b \to s \bar d d$ and/or $b \to s \bar u u$, and some involve mixtures.
Moreover, QCD corrections can differ for different final states.  The
experimental values have shifted a good deal from year to year, providing
theorists with a moving target which they have been quite adept at following.
At present the number on which I am keeping an eye is that from $B^0 \to \pi^0
K_S$, which both BaBar and Belle agree lies below the reference value, with an
average $\sin 2 \beta_{\rm eff} = 0.33 \pm 0.21$.  (Note the large experimental
error.)  The value of the $\cos \Delta m t$ coefficient $C_{K_S \pi^0} = 0.12
\pm 0.11$ also is interesting.  This is just $-A_{CP}(K^0 \pi^0)$.  As noted
earlier, sum rules predict a central value of 0.14 to 0.15 for $C_{K_S \pi^0}$.

Many estimates have been performed of deviations of $\sin 2 \beta_{\rm eff}$
from the reference value in the Standard Model.  Typical explicit calculations
give a deviation of 0.05 or less, usually predicting $\sin 2 \beta_{\rm eff}$
{\it larger} than 0.678 whereas most experiments find {\it lower} values.
Flavor-SU(3) estimates~\cite{sin2bdiff} allow differences of at most 0.1.
 
\subsection{$CP$ violation in $B \to \pi \pi$}

An example of the systematic error associated with uncertainty in hadron
physics is provided by a detailed examination of time-dependent CP asymmetries
in $B^0 \to \pi^+ \pi^-$.  This is relevant to remarks made by L\"u
\cite{Lu} at this Conference concerning limitations in our ability to
learn the weak phases $\alpha$ and $\gamma$.  I report on work
with M. Gronau,~\cite{MGJR07} updating a previous analysis.~\cite{MGJR04}

\begin{figure}
\begin{center}
\includegraphics[height=3.9in]{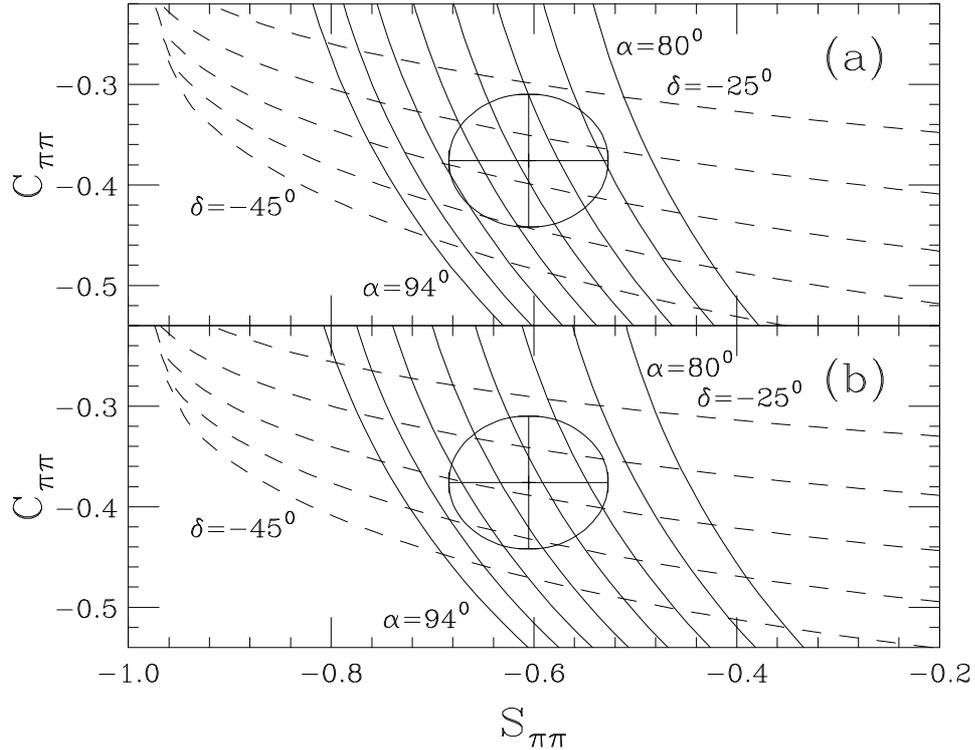}
\end{center}
\caption{Values of $C_{\pi \pi}$ plotted against $S_{\pi \pi}$ for values of
$\alpha$ spaced by 2 degrees (solid curves) and $\delta$ spaced by 5 degrees
(dashed contours).  The degree of penguin ``pollution'' is estimated in (a)
from $B^+ \to K^0 \pi^+$ and in (b) from $B^0 \to K^+ \pi^-$.
\label{fig:cs}}
\end{figure}

The time-dependent asymmetry parameters $(S_{\pi \pi}, C_{\pi \pi})$ have been
measured by BaBar \cite{BaSC} ($-0.60 \pm 0.11,-0.21 \pm 0.09$) and Belle
\cite{BeSC} ($-0.61 \pm 0.11,-0.55\pm 0.09$), leading to an average~\cite{hfag}
$(-0.605 \pm 0.078,-0.376 \pm 0.066)$.  These average values are plotted in
Fig.\ \ref{fig:cs} along with predictions for values of the weak phase $\alpha$ 
and strong phase $\delta = \delta^P - \delta^T$.  An SU(3)-breaking factor
$f_K/f_\pi = 1.22$ has been taken for the ratio of $|\Delta S| = 1$ to
$\Delta S = 0$ tree amplitudes, but no SU(3) breaking has been assumed for
the corresponding ratio of penguin amplitudes.  The error ellipses represented
by the plotted points encompass the ranges $81^\circ \le \alpha \le 91^\circ$
(implying $68^\circ \le \gamma \le 78^\circ$) and $-40^\circ \le \delta \le
-26^\circ$.  As in Ref.\ \cite{Fleischer:2007}, we get a very small range
of $\gamma$ [here $(73 \pm 4)^\circ$], but additional systematic errors are
important.  In the upper figure, the penguin ``pollution'' has been estimated
using $B^+ \to K^0 \pi^+$, entailing the neglect of a small ``annihilation''
amplitude, while in the lower figure it has been estimated using $B^0 \to K^+
\pi^-$. in which the effect of a small tree amplitude must be included.  The
two methods give weak phases within a degree or two of one another.

Now we examine the effect of SU(3) breaking in the ratio of penguin amplitudes.
Call the $\Delta S = 0$ penguin $P$, the $|\Delta S| = 1$ penguin $P'$, and
define $\xi_P \equiv |P'/P|V^*_{cd}V_{cb}/V^*_{cs}V_{cb}|$.  The above
exercise was for $\xi_P = 1$.  Now we vary $\xi_P$.

One could assume $\xi_P = f_K/f_\pi = 1.22$ as for the tree amplitude ratio.%
~\cite{Fleischer:2007}  Alternatively, one could determine it from $\Delta(K^+
\pi^-) = - \xi_P \Delta(\pi^+ \pi^-)$, where $\Delta(f) \equiv \Gamma(\bar B
\to \bar f) - \Gamma(B \to f)$.  In this case with the world average
$A_{CP}(K^+ \pi^-) = -0.097 \pm 0.012$ one finds $\xi_P = 0.79 \pm 0.18$.

The change from $\xi_P= 1$ to $\xi_P = 1.22$ shifts $\alpha$ up ($\gamma$ down)
by $\sim 8^\circ$, $|\delta|$ up by $\sim 10^\circ$, while the change to $\xi_P
= 0.79$ shifts $\alpha$ down ($\gamma$ up) by $\sim 10^\circ$, $|\delta|$
down by $\sim 8^\circ$.  The systematic (theory) errors are larger than
the statistical ones.  As stressed by L\"u,~\cite{Lu} one needs to gain control
of SU(3) breaking.  In order to provide information beyond that obtained from
flavor SU(3), schemes such as PQCD \cite{Lu} and SCET \cite{Becher}
need to predict $\delta$ to better than $10^\circ$.

Discussion at this Conference concerned the relative merits of
frequentist \cite{freq} and Bayesian~\cite{Bay} analysis, referring to a
recent controversy over what can be learned from $B \to \pi
\pi$.~\cite{deedum}  The intelligent choice of priors can have merits, e.g.,
when searching for a point on the surface of a sphere (taking a uniform prior
in the cosine of the polar angle $\theta$, not $\theta$ itself) or when
searching for a lost skier at La Thuile (beginning by looking near the lifts).

\section{$D$ mixing}

In the Standard Model, mixing due to shared intermediate states reached by
$|\Delta C| = 1$ transitions dominates $D^0$--$\od$ mixing. In the flavor-SU(3)
limit these contributions (e.g., $\pi \pi$, $K \bar K$, $K \pi$, and $\bar K
\pi$) cancel one another.~\cite{canc}  How precise is the cancellation?

Define $D_1$ and $D_2$ to be the mass eigenstates (respectively CP-even and
-odd in the absence of CP violation), $\Delta M \equiv M_1 - M_2$, $\Delta
\Gamma \equiv \Gamma_1 - \Gamma_2$, $x \equiv \Delta M/\Gamma$, and $y \equiv
\Delta \Gamma/ \Gamma$, where $\Gamma \equiv (\Gamma_1 + \Gamma_2)/2$.
Estimates of $y$ range up to ${\cal O}(1\%)$, with $|x| \le |y|$ typically.

The time dependence of ``wrong-sign'' $D^0(t=0)$ decays (e.g., to $K^+ \pi^-$)
involves the combinations $x' \equiv x \cos \delta_{K \pi} + y \sin \delta_{K
\pi}$, $y' \equiv - x \sin \delta_{K \pi} + y \cos \delta_{K \pi}$, where the
strong phase $\delta_{K \pi}$ has been measured by the CLEO Collaboration:%
~\cite{CLEOdKpi} $\cos \delta_{K \pi} = 1.09 \pm 0.66$.  In the SU(3) limit,
$\delta_{K \pi} = 0$.~\cite{nodKpi}  This method has been used by the BaBar
Collaboration \cite{Flood,Bamix} to obtain the non-zero mixing parameter
$y' = (9.7 \pm 4.4 \pm 3.1) \times 10^{-3}$.

The Belle Collaboration has obtained evidence for mixing in a different way, by
comparing lifetimes in CP- and flavor-eigenstates and thereby measuring a
parameter $y_{CP} = (1.13 \pm 0.32 \pm 0.25)\%$.~\cite{Belife,Staric}  In the
limit of CP conservation (a likely approximation for $D$ mesons), $y_{CP} = y$.
A time-dependent Dalitz plot analysis of $D^0 \to K_S \pi^+ \pi^-$ by Belle
\cite{Staric,Bemix} obtains $x = (0.80 \pm 0.29^{+0.09+0.15}_{-0.04-0.14})\%$,
$y = (0.33 \pm 0.24^{+0.07+0.08}_{-0.12-0.09})\%$.

These results were synthesized in several theoretical analyses.~\cite{mixth}
The consensus is that while $y$ is near the upper limit of what was anticipated
in the Standard Model, there is no evidence for new physics.  Observation of
CP violation in $D$ decays, on the other hand, would be good evidence for such
physics, and will continue to be the object of searches.

\section{Low-energy hadron physics}

\begin{figure}
\mbox{\includegraphics[height=1.95in]{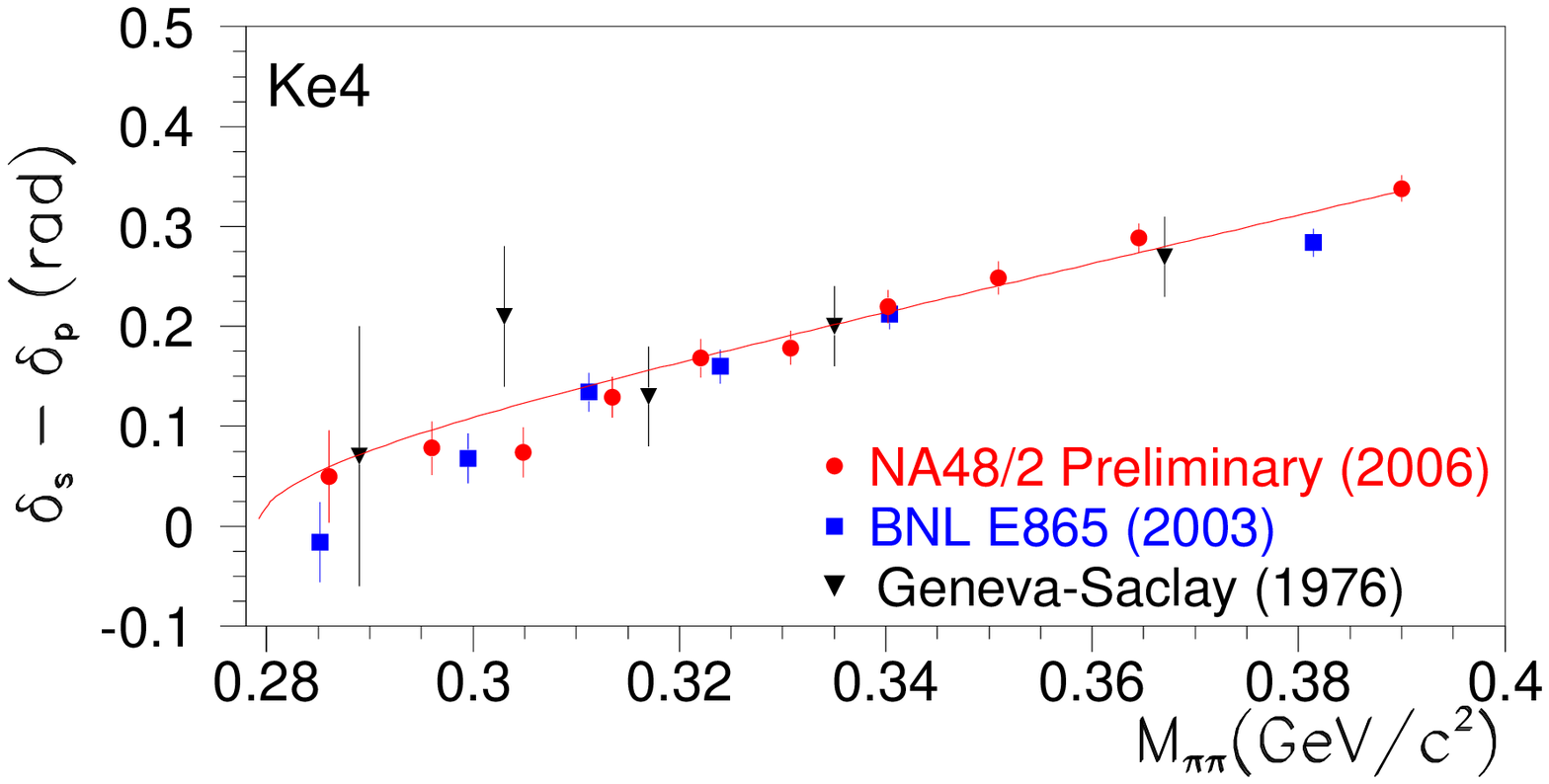}
      \includegraphics[height=1.95in]{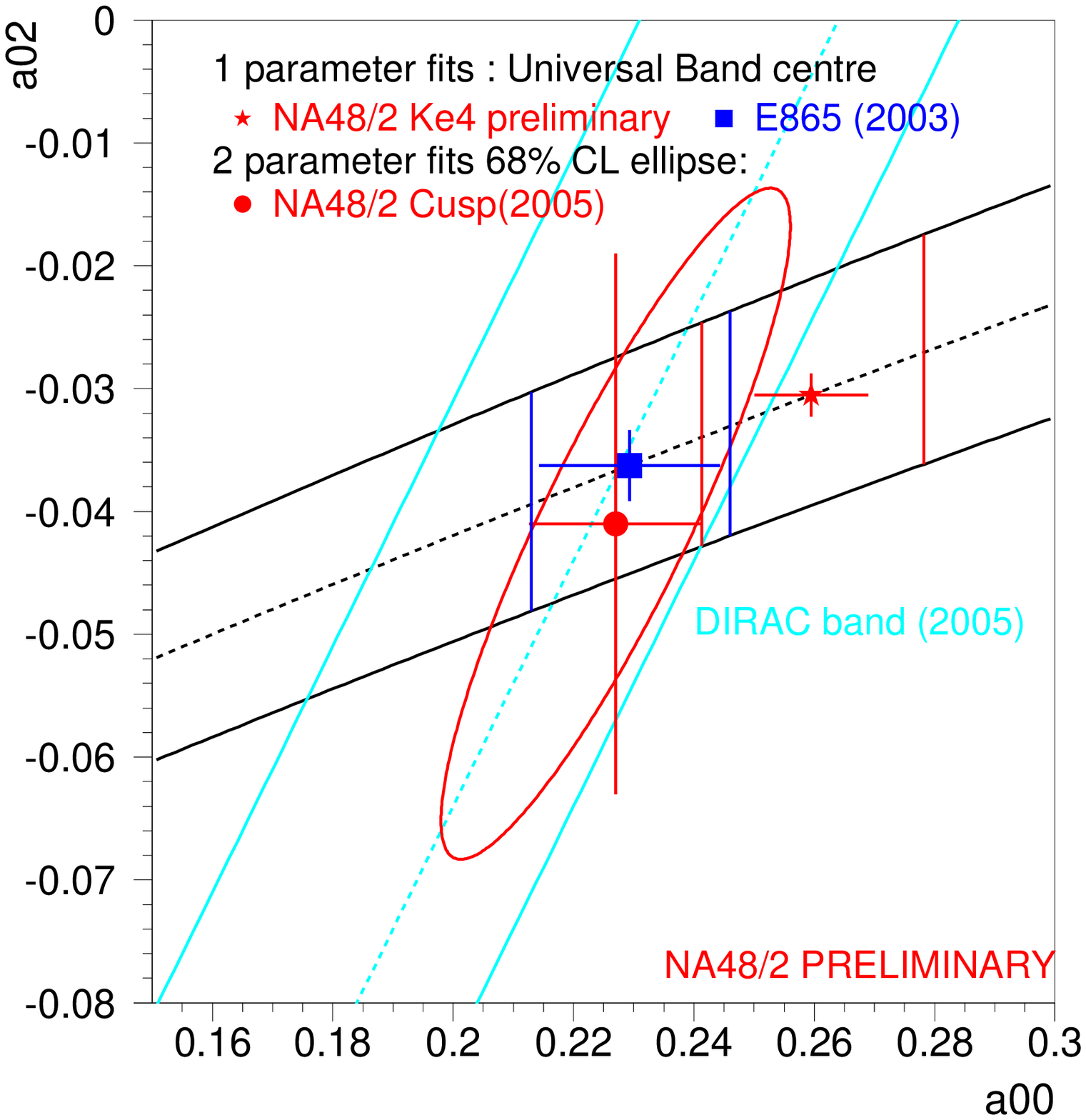}}
\caption{Information on $\pi \pi$ scattering from NA48 and other sources.%
\protect \cite{NA48}  Left:  $K_{e4}$ decays; right: $\pi \pi$ scattering
lengths.
\label{fig:NA48}}
\end{figure}

Information on light-quark interactions and spectroscopy continues to
accumulate from weak decays of kaons, charm (telling about the low-mass $I=J=0$
dipion resonance $\sigma$), and $B$ (illuminating properties of scalar
mesons like $f_0$ and $a_0$, which must be understood if one is to identify
glueballs), and radiative $\phi$ decays.  For example, the
NA48 Collaboration at CERN has obtained information on $\pi \pi$ scattering
lengths from $K_{e4}$ and $K^+ \to \pi^+ \pi^0 \pi^0$ decays.~\cite{NA48}
Some results are summarized in Fig.\ \ref{fig:NA48}.

Scattering lengths $a_J^I$ are conventionally labelled by total momentum $J$
and isospin $I$.  The predictions of current algebra \cite{Leut} are $a_0^2 =
-0.044$ and $a_0^0 = 0.22$.  The NA48 measurement of $a_0^0$ seems to be
slightly above this last value but more data from NA48 will tell whether there
really is a discrepancy.

The helicity structure of $\rho$ mesons in the reaction $e^+ e^- \to \rho^+
\rho^-$ has  recently been measured by the BaBar Collaboration,~\cite{Blanc}
with the result $F_{00} = 0.54 \pm 0.10 \pm 0.02$, where the subscripts denote
$\rho$ helicity.  This is to be compared with the asymptotic prediction
\cite{BL} $F_{00} \to 1$.  Should one be surprised?  Are there related tests
at comparable scales of $E_{\rm cm} \simeq 10$ GeV?

Recent results by the KLOE Collaboration \cite{Ambrosino:2006gk,DiMicco} shed
light on the quark/gluon content of $\eta'$ through the decay $\phi \to \eta'
\gamma$.  Comparison of this decay with others (such as $\phi \to \eta \gamma$,
$\rho \to \eta \gamma$, $\eta \to \gamma \gamma$, $\eta' \to \gamma \gamma$,
and so on), following a method proposed some time ago,~\cite{JR83} lead to the
conclusion that the glue content of the ${\eta'}$ is (14$\pm$4)\%.

\section{Charmonium}

Results from BES were presented at this Conference \cite{Wang,Ma} on states
reached in $J/\psi$ decays, including a broad $X(1580)$ decaying to $K^+ K^-$
seen in $J/\psi \to K^+ K^- \pi^0$ and an $\omega \phi$ threshold peak seen
in $J/\psi \to \gamma \omega \phi$, as well as on multibody $\psi(2S)$ decays.
CLEO results \cite{Dubrovin,Mitchell} included confirmation of the $Y(4260)$
in a direct scan and in radiative return; a new measurement of $M(D^0)$ which
implies that the $X(3872)$ is bound by $0.6 \pm 0.6$ MeV; and observation of
$\psi''(3770) \to \gamma \chi_c$ decays with rates confirming its assignment as
the $1^3D_1$ charmonium state.  Belle \cite{Uehara} reported two-photon
production of several states including $Z(3930)$, a $\chi_{c2}(2P)$
candidate.

\section{Charmed hadrons}

\subsection{$L=0$ states}

BaBar \cite{Aubert:2006je,Zhang} has identified the $\Omega_c^*$, a candidate
for the lowest-lying $J=3/2$ $css$ state lying $70.8 \pm 1.0 \pm 1.1$ MeV above
the $\Omega_c$ (also recently studied by BaBar~\cite{Aubert:2007bt}).  This
mass splitting agrees with that predicted in the quark
model.~\cite{JROm}  One now has a complete set of candidates for the $L=0$
mesons and baryons containing a single charmed quark.  As we shall see, charmed
hadron masses are useful in anticipating those of hadrons containing a $b$
quark.

\subsection{Orbitally-excited mesons}

In the heavy-quark limit, mesons made of one heavy and one light quark are best
described by coupling the light quark and the orbital angular momentum $L$ to
a total $j$, and then $j$ to the heavy quark spin to form states of $J = j \pm
1/2$.  For $L=1$ one then has states with $j = 1/2$ (leading to $J=0,1$) and
$j=3/2$ (leading to $J=1,2$).  The $J=3/2$ states, predicted to be narrow,
have been known for many years for both charmed-nonstrange and charmed-strange
mesons. However, the $j=1/2$ states, expected to be broad, proved more elusive.

The two $L=1,~j=1/2$ $c \bar s$ mesons, the $D_{s0}(2317)$ and $D_{s1}(2460)$,
were lighter than expected by most theorists.  Lying below the respective $DK$
and $D^*K$ thresholds for strong decays, they turned out to be narrow, decaying
radiatively or via isospin-violating $\pi^0$ emission.  Their low masses
{\it were} anticipated in schemes which pegged them as chiral partners of
the $D_s$ and $D^*_s$.~\cite{chiral}  Regarding them as bound states
of $DK$ and $D^*K$, respectively, they each would have a binding energy of
41 MeV.  It would be interesting to see if a similar pattern holds for $B_{sJ}$
as $\bar B^{(*)}K$ bound states.  The lesson is that light-quark degrees of
freedom appear to be important in understanding heavy-quark systems.

Higher-mass $c \bar s$ states have now been reported.~\cite{Grenier}  The Belle
Collaboration \cite{Abe:2006xm} sees a $D_s$ state in the $M(D^0 K^+)$ spectrum
in $B^0 \to \bar D^0 D^0 K^+$.  It has $M = (2715 \pm 11^{+11}_{-14})$ MeV and
$\Gamma = (115 \pm 20^{+36}_{-32})$ MeV.  BaBar could be seeing this state,
though not with significance.~\cite{Aubert:2006mh}  It has $J^P = 1^-$ and lies
$603^{+16}_{-18}$ MeV above $D_s^*(2112)$, to be compared with 2S--1S
splittings of 681$\pm$20 MeV for $s \bar s$ and 589 MeV for $c \bar c$.
It appears to be a good $c \bar s(2^3S_1)$ candidate.

Another $D_s$ state is seen decaying to $D^0 K^+$ and $D^+ K_S$.%
~\cite{Aubert:2006mh}
It has $M = (2856.6 \pm 1.5 \pm 5.0)$ MeV and $\Gamma = (48 \pm 7 \pm 10)$ MeV.
It can be interpreted as the first radial excitation of $D_{s0}(2317)$
\cite{radex2857} or a $J^P = 3^-(^3D_3)$ state.~\cite{3D2857}  Angular
distributions of decay products should permit a distinction.

While the established (narrow) $j^P = 3/2^+$ states $D_1(2422),~D_2(2460)$ have
been known for quite some time, there is more question about the broad $j^P =
1/2^+$ candidates.  Both CLEO~\cite{CLEO1} and Belle \cite{Belle01} place the
broad $j^P = 1/2^+,~J^P = 1^+$ candidates in the narrow range 2420--2460 MeV,
but Belle \cite{Belle01} and FOCUS \cite{FOCUS0} differ somewhat with respect to
broad $j^P = 1/2^+,~J^P = 0^+$ candidates, placing them only in a rather wide
range 2300--2400 MeV.

One feature of note is that orbital excitation to the well-established $j=3/2$
states costs (472,482) MeV for ($D^{**},D_s^{**}$).  We shall compare this
figure with a corresponding one for $B$ mesons.

\section{Beauty hadrons}

\subsection{$L=0$ states}

CDF has observed $\Sigma_b$ and $\Sigma^*_b$ candidates decaying to $\pi^\pm
\Lambda_b^0$.~\cite{CDFsigb,Filthaut} Their mass measurements are aided by a new
precise value, also due to CDF,~\cite{Acosta:2005mq} $M(\Lambda_b) = (5619.7
\pm 1.2 \pm 1.2)$ MeV.  It is worth comparing this mass with a simple quark
model prediction.

The light ($u,d$) quarks in $\Lambda_c$ and $\Lambda_b$ must be coupled to spin
zero, by the requirements of Fermi statistics, as they are antisymmetric in
color ($3^*$) and flavor ($I=0$) and symmetric in space (S-wave).  Aside from
small binding effects, one then expects $M(\Lambda_b) - M(\Lambda_c) = M_b -
M_c$, where $M_b$ and $M_c$ are ``constituent'' quark masses whose difference
$M_b - M_c$ may be obtained from $B^{(*)}$ and $D^{(*)}$ mesons by taking the
combinations $(3 M^* + M)/4$ for which the hyperfine $Q \bar q$ interactions
cancel.  Using $[3 M(B^*) + M(B)]/4 = 5314.6 \pm 0.5$ MeV and $[3M(D^*) + M(D)]
/4 = 1973.0 \pm 0.4$ MeV one then finds $M_b - M_c = 3341.6 \pm 0.6$ MeV.
(This is slightly larger than the difference between $M_b=4796$ MeV and $M_c=
1666$ MeV reported by K\"uhn.~\cite{Kuhn})  Combining this difference with
$M(\Lambda_c) = 2286.46 \pm 0.14$ MeV, one then predicts $M(\Lambda_b) = 5628.1
\pm 0.7$ MeV, 8 MeV above the observed value.  One could ascribe the small
difference, which goes in the right direction, to reduced-mass effects.  A
similar exercise predicts $M(\Xi_b) \simeq 5.8$ GeV from $M(\Xi_c) = 2469$ MeV.

We now turn to the $\Sigma_b^{(*)}$ states.  The direct measurements are of
$Q^{(*)\pm} \equiv M(\Sigma_b^{(*)\pm}) - M(\pi^\pm) - M(\Lambda_b)$, and it is
found (under the assumption $Q^{*+} - Q^{*-} = Q^+ - Q^-$, which is expected
to be good to 0.4 MeV \cite{Rosner:2006yk}) that
\beq
Q^+ = 48.4^{~+~2.0~+~0.2}_{~-~2.3~-~0.3}~{\rm MeV}~,~~
Q^- = 55.9 \pm 1.0 \pm 0.2 ~{\rm MeV}~.
\eeq
With the new CDF value of $M(\Lambda_b)$, these results then imply
\bea
M(\Sigma_b^-) & = & 5815.2^{~+~1.0}_{~-~0.9}\pm 1.7~{\rm MeV}~,~~
M(\Sigma_b^+) = 5807.5^{~+~1.9}_{~-~2.2} \pm 1.7~{\rm MeV}~,\\
M(\Sigma_b^{*-}) & = & 5836.7^{~+~2.0~+1.8}_{~-~2.3~-~1.7}~{\rm MeV}~,~~
M(\Sigma_b^{*+}) = 5829.0^{~+~1.6~+1.7}_{~-~1.7~-1.8}~{\rm MeV}~.
\eea

These masses are entirely consistent with quark model predictions.  (See
\cite{Rosner:2006yk} and references therein.)  The $\Lambda$ hyperon may be
denoted $[ud]s$, where $[ud]$ denotes a pair antisymmetric in flavor and spin,
whereas the $\Sigma^{+,0,-}$ quark wavefunction may be written as
$(\cdot \cdot)s$, with $(\cdot\cdot) = (uu), (ud), (dd)$ shorthand for a pair
{\it symmetric} in flavor and spin.  $S(\cdot \cdot) = 1$ then can couple with
$S(s) = 1/2$ to give $J = 1/2~(\Sigma)$ or 3/2 ($\Sigma^*$), with hyperfine
splitting $\propto 1/m_s$. The mass difference between the spin-1 and
spin-0 diquarks, $M(\cdot \cdot) - M[ud] = [2 M(\Sigma^*) + M(\Sigma)]/3
- M(\Lambda)$, can be calculated from the spin-weighted average of
$M(\Sigma^*)$ and $M(\Sigma)$, in which hyperfine interactions cancel out.
This result is the same calculated from baryons containing $s$, $c$, or $b$:
\beq
\frac{\Sigma + 2 \Sigma^*}{3} - \Lambda = 205.1 \pm 0.3 {\rm~MeV}~.
\frac{\Sigma_c + 2 \Sigma_c^*}{3} - \Lambda_c = 210.0 \pm 0.5 {\rm~MeV}~,~~
\eeq
to be compared with
\beq
\frac{\Sigma_b + 2 \Sigma_b^*}{3} - \Lambda_b = 205.9 \pm 1.8 {\rm~MeV}~.
\eeq
The hyperfine splittings themselves also obey reasonable scaling laws.  One
expects $M(\Sigma^*) - M(\Sigma) \propto 1/m_s$ so splittings for charm and
bottom should scale as $1/m_c,~1/m_b$, respectively.  The differences for
$s$, $c$, and $b$, $191.4 \pm 0.4,~64.4 \pm 0.8,~21.3 \pm 2.0$ MeV, are indeed
approximately in the ratio of $1/m_s~:~1/m_c~:~1/m_b$.

\subsection{$L=1$ mesons}

Results from CDF and D0, summarized by Filthaut,~\cite{Filthaut} are shown in
Table \ref{tab:Bx}.  Arguments similar to those for the $L=0$ baryons in the
previous subsection imply that one should expect $M(B_2) - M(B_1) \simeq
M(B_{s2}) - M(B_{s1}) \simeq 13$ MeV.  This pattern does not seem to emerge
clearly from the data, which in any case give mixed signals regarding
hyperfine splittings.  One pattern which does seem fairly clear is that
orbital $j=3/2$ $B,B_s$ excitations cost $\sim 50$ MeV less than for $D,D_s$.

\begin{table}
\caption{Candidates for $L=1,~j^P=3/2^+$ $B$ mesons.  Masses in MeV.
\label{tab:Bx}}
\begin{center}
\begin{tabular}{c c c c c} \hline
 & \multicolumn{2}{c}{Nonstrange} & \multicolumn{2}{c}{Strange} \\ \hline
    & $B_1$ & $B_2$ & $B_{s1}$ & $B_{s2}$ \\ \hline
CDF & 5738$\pm$5$\pm$1 & 5734$\pm$3$\pm$2 & 
 5829.2$\pm$0.2$\pm$0.6 & 5839.6$\pm$0.4$\pm$0.5 \\
D0  & 5720.8$\pm$2.5$\pm$5.3 & 5746$\pm$2.4$\pm$5.4 &
 -- & 5839.1$\pm$1.4$\pm$1.5 \\ \hline
\end{tabular}
\end{center}
\end{table}

\section{Importance of thresholds}

Many hadrons discovered recently require that one understand nearby thresholds,
a problem with a long history.~\cite{Fano,Wigner,Feshbach}  As one example,
the cross section for $e^+ e^- \to$ (hadrons) has a sharp dip around a
center-of-mass energy of 4.25 GeV, which is just below the threshold for the
lowest-lying pair of charmed mesons ($D^0$ and $\bar D_1^{*0}$) which can be
produced {\it in a relative S-wave}.  All lower-mass thresholds, such as
$D \bar D$, $D \bar D^*$, and $D^* \bar D^*$, correspond to production in
relative P-waves, so the corresponding channels do not open up as quickly.
The $D^0 \bar D_1^{*0}$ (+ c.c.) channel is the expected decay of the
puzzling charmonium state $Y(4260)$ if it is a hybrid ($c \bar c$ + gluon).
But this channel is closed, so others (such as the observed $\pi \pi J/\psi$
channel) may be favored instead.

It is likely that the dip in $e^+ e^- \to$ (hadrons) is correlated with a
substantial suppression of charm production just before the $D^0 \bar D_1^{*0}$
channel opens up.  The cross section for $e^+ e^- \to D^* \bar D^*$ (a major
charm channel) indeed experiences a sharp dip at 4.25 GeV.~\cite{Bedip}
Perhaps the peak $Y(4320) \to \pi^+ \pi^- \psi(2S)$ seen by BaBar,~\cite{Ba2S}
with $M = 4324 \pm 24$ MeV, $\Gamma = 172 \pm 33$ MeV, is correlated with some
other threshold.

Many other dips are correlated with thresholds [e.g., in the $\pi \pi$ S-wave
near $2M(K)$ or $\gamma^* \to 6 \pi$ near $2 M(p)$.~\cite{JRth}]  The BaBar
Collaboration recently has reported a structure in $e^+ e^- \to \phi
f_0(980)$ at 2175 MeV.~\cite{Bath}  It could be a hybrid $s \bar s g$
candidate in the same way that $Y(4260)$ is a hybrid $c \bar c g$ candidate.
The assignment makes sense if $M_c - M_s \simeq (M_Y - M_X)/2 = 1.04$ GeV.

\section{Quark masses}

J. H. K\"uhn \cite{Kuhn} has presented explicit formulae for the running of
quark masses.  High-order corrections to the Taylor series for the heavy quark
vacuum polarization function $\Pi_Q(q^2)$ are a {\it tour de force}.  [One may
expect interesting things from this group on high-order corrections to $R =
\sigma(e^+e^- \to {\rm hadrons})/\sigma(e^+e^- \to \mu^+ \mu^-)$.] The moments
${\cal M}_n = \int ds R(s)/s^{n+1}$
give consistent masses, with $m_c(m_c) = 1287 \pm 13$ MeV from $n=1$ and
$m_b(m_b) = 4167 \pm 23$ MeV from $n=2$.  These results are an update of
Ref.\ \cite{moments}.  The pole masses $M_b = 4796$ MeV and $M_c = 1666$ MeV
differ by 3130 MeV, a bit less than the phenomenological value of 3342 MeV
mentioned earlier in the prediction of $M(\Lambda_b)$.  One caveat is that old
CLEO data were used with an arbitrary renormalization.  CLEO should come out
soon with new $R$ values below $B \bar B$ threshold but needs to present
its data above $B \bar B$ threshold similarly.  These data were taken
in connection with a search for $\Lambda_b \bar \Lambda_b$ production.%
~\cite{CLEOLLb}

A. Pineda has reminded me of a work \cite{PS} in which $\bar m_b(\bar m_b) =
4.19 \pm 0.06$ GeV is obtained from a non-relativistic sum rule.  K\"uhn's talk
has a compilation of many other values.  The uncertainty in $m_c$, reduced by
K\"uhn's analysis, is an important part of the theoretical error in
calculating ${\cal B}(b \to s \gamma)$.~\cite{Haisch}

Although the top quark mass has been measured with impressive accuracy (see
below), it may be possible by studying threshold behavior in $e^+ e^-
\to t \bar t$ to learn it to about 0.1 GeV.~\cite{Penin}

\section{Heavy flavor production}

Calculations of hadronic charm production are in rough accord with experiment
(though there remains some excess peaking for small azimuthal angle between
charm and anticharm).  While the description of beauty production has improved
vastly in the past few years, there are still some kinematic regions where
experiment exceeds theory.~\cite{HQprod}  Incisive beauty--antibeauty
correlation measurements still do not exist despite long-standing pleas.%
~\cite{Kwong89}  One looks forward to these at the LHC.~\cite{Andreev}

The quantitative understanding of quarkonium production still seems elusive.
It demands soft gluon radiation, ``adjustable'' to the observed
cross section.  This is not the same as a first-principles calculation.

\section{Fragmentation and jets}

The correct description of fragmentation was a key ingredient in improving
the agreement of $b$ production predictions with experiment.~\cite{HQprod}  At
this conference new and/or upgraded Monte Carlo routines were reported.%
~\cite{Krauss,Richardson}  A useful detailed check of their hadronization
features would be to compare their predicted multiplicities and particle
particle species with CLEO data on hadronic $\chi_c$ decays \cite{Mitchell} or
hadronic bottomonium decays (which are being analyzed by CLEO).  One could also
imagine applying the global determination of fragmentation functions reported
by Kumano \cite{Kumano} to these questions.

Progress also has been reported with spinor-based multigluon methods;%
~\cite{Varman,Badger} definition of $b$-jets;~\cite{Zanderighi} correction for
the underlying event;~\cite{Cacciari} exclusive $p \bar p \to p \bar p X$
reactions;~\cite{Terashi} inclusive cross sections;~\cite{Cwiok,Norniella} and
an infrared-safe-safe jet definition.~\cite{Salam}  Jets in heavy-ion collisons
will be especially challenging.~\cite{jetsHI}

\section{$W$ and top}

New CDF values of $(M_W = 80413 \pm 48)$ MeV and $\Gamma_W = (2032 \pm 71)$ MeV
have recently been reported.~\cite{WMG}  The new world averages, $M_W = 80398
\pm 25$ MeV and $\Gamma_W = (2095 \pm 47)$ MeV, are consistent with the
Standard Model.  In the latter there is very little room for deviations since
no ``oblique'' ($S,T$) corrections are expected:~\cite{GW}
\beq
\Gamma(W) = \frac{G_\mu M_W^3}{6 \pi \sqrt{2}} \left\{ 3 + 6 \left[1 +
\frac{\alpha_S(M_W)}{\pi} \right] \right\} = (2100 \pm 4)~{\rm MeV}~.
\eeq

Now information on top quark mass and production comes from CDF and D0.%
~\cite{top}  Examples of new measurements in the $\ell$ + jets channel are
$m_t = (170.5 \pm 2.4 \pm 1.2)$ GeV (D0) and $(170.9 \pm 2.2 \pm 1.40)$ GeV
(CDF).  The present world average is now $m_t = (170.9 \pm 1.8)$ GeV, an
error of 1.1\%.  This places further pressure on the Higgs mass.  The Standard
Model fit gives $M_H \le 144$ GeV (95\% c.l.), relaxed to 182 GeV
if the present direct limit $M_H > 114.4$ GeV is considered.

One alternative to a light Higgs boson would involve custodial symmetry
violation [for example, as provided by a new heavy SU(2) doublet with large
mass splitting].~\cite{PW}  Adding a vacuum expectation value $\langle V_0
\rangle$ of a Higgs triplet with zero hypercharge which is only a few percent
of the standard doublet $v=246$ GeV would be sufficient to subsantially relax
the upper limit on $M_H$.~\cite{JRHiggs}

The D0 Collaboration sees single-top production at the expected level in
three different analyses.~\cite{D0top}  CDF sees it in one analysis but not in
two others.~\cite{CDFtop}  When the dust settles, this measurement is expected
to provide useful information on $|V_{tb}|$.

\section{Dibosons and Higgs}

CDF and D0 have presented evidence for $WZ$ and $ZZ$ production, as summarized
by F. W\"urthwein.~\cite{Wurthwein}  D0 has seen a dip corresponding to the
expected radiation zero in $W \gamma$ production.  The subprocess $u \bar d \to
W^+ \gamma$ has a zero at $\cos \theta_{\rm CM} = -1/3$, while $\bar u d \to
W^- \gamma$ has a zero at $\cos \theta_{\rm CM} = 1/3$.

In a search for the Higgs boson in the $H \to \tau \tau$ channel, bounds from
CDF are ``degraded'' thanks to an excess of events for $M_H \simeq 160$ GeV.
On the other hand, D0 sees a deficit there.~\cite{Higgs}  This mass range may
be the first interval accessible with 8 fb$^{-1}$ at the Tevatron;
sensitivities are improving faster than 1/$\sqrt{\int {\cal L} dt}$.%
~\cite{Kilminster}  It would
be wonderful if a way were found to extend the run!

An interesting scheme for generating the Higgs boson via spontaneous conformal
symmetry breaking was presented.~\cite{Meissner}  As this tends to give a
fairly heavy Higgs boson, it must be confronted with the tightening precision
electroweak constraints.  Strong electroweak symmetry breaking scenarios also
were described.~\cite{Allwood}  These essentially adapt chiral models to the
TeV scale, replaying the strong interactions at a factor $v/f_\pi \simeq 2650$
higher in energy.  Light-Higgs scenarios are not ruled out; for instance,
it has been asked whether the mass of the $b \bar b(1^1S_0)$ state, the
as-yet-unseen $\eta_b$, is standard or is affected by mixing with a light
Higgs boson.~\cite{SL}  One Standard Model prediction
\cite{Kniehl:2003ap,Pineda} is $M(\eta_b) = 9421$ MeV.

Higgs decays to multiparticle final states have been described using twistor
methods.~\cite{Badger}  It may be possible to produce a Higgs boson at LHC in
the double-diffractive reaction $p p \to p p H$, monitoring the small-angle
protons using Roman pots.~\cite{Terashi}  One problem will be distinguishing
which of the multiple interactions per crossing was the source of the
scattered protons.  This pileup effect may be soluble if one can make
sufficiently rapid trigger decisions.

Two-Higgs models, if confirmed, provide a gateway to supersymmetry.%
~\cite{Weiglein}  Such proliferation of the Higgs spectrum, entailing two
charged and three neutral Higgs bosons, also is a feature of grand unified
theories beyond the minimal SU(5), such as SO(10).

\section{Proton structure and diffraction}

The proton spin $\frac{1}{2}$ is composed of $\frac{1}{2} \Delta \Sigma
+ \Delta G + \Delta L$, corresponding respectively to quarks, gluons, and
orbital angular momentum.  $\Delta \Sigma \simeq 0.3$; what's the rest?
The COMPASS \cite{compass} and STAR \cite{star} Collaborations have shown that
$\Delta G$ is not enough; one must have $\Delta L > 0$.

Neutral-current $ep$ interactions at HERA have displayed the first evidence
for parity violation in high-$Q^2$ deep inelastic scattering.~\cite{HERAPV}
HERA is helping to pin down structure functions and their evolution for use at
the LHC.~\cite{HERAsf}  Also at HERA, it has been found that the Pomeron slope
is different in $\rho^0$ and $J/\psi$ photoproduction.  These reactions
correspond respectively to soft and hard processes.~\cite{HERApp}

\section{Heavy ion collisions}

One has seen the adaptation of string theory ideas to properties of the
quark-gluon plasma: hydrodynamic properties involve previously intractable
strong-coupling calculations.~\cite{JCS}  In heavy-ion jet production, the
recoiling jet is quenched if it must pass through the whole nucleus.%
~\cite{jetsHI}  This provides information about the properties of nuclear
matter.  An interesting rapidity ``ridge'' is seen in many processes.  Could
this be a manifestation of QCD ``synchrotron radiation''?  Do previous emulsion
experiments~\cite{Deines-Jones:1996fe} display this feature?

One way to describe nuclear matter effects is via medium-modified fragmentation
functions probe nuclear matter effects.~\cite{Cunqueiro,Belghobsi}  Useful
information is provided by $\gamma$--$\pi^0$ and $\gamma$--$\gamma$
correlations.~\cite{Belghobsi}  Hanbury-Brown-Twiss correlations between
identical particles (e.g., $\pi^\pm \pi^\pm$) provide information on the
viscosity of the quark-gluon plasma and on the geometry and time evolution of
the ``hot'' region.~\cite{HBT}

Charmed particles are found to interact with the nuclear medium in the same way
as others.~\cite{charmint}  It is not clear whether there is a difference
between the interactions of $c \bar q$ and $\bar c q$ states; certainly
$K^+$ and $K^-$ do interact differently with nonstrange matter.
Other important issues in nuclei include low-$x$ parton saturation
\cite{partsat} and the question of whether quarkonium suppression is taking
place.~\cite{qsupp}

\section{Beyond the Standard Model}

As this is a large field, I would like to comment on just a few items which I
consider especially worth watching in the next few years.

(1) The muon's $g-2$ value can get big contributions in some SUSY models.
In units of $10^{-11}$, $a_\mu \equiv (g_\mu-2)/2 = 116~591~793~(68)$ (theory),
to be compared with 116~592~080~(63) (experiment).  These differ by $(287 \pm
93)$ or $3.1 \sigma$.~\cite{Jegerlehner}  This relies upon evaluating hadronic
vacuum polarization via $e^+ e^-$ annihilation. If one uses $\tau$ decays the
discrepancy drops to $1.2 \sigma$.  The inconsistency is worth sorting out.

(2) Non-standard explanations abound for the deviation of the effective $\sin(2
\beta)$ in $b \to s$ penguins from the ``reference value'' obtained in decays
dominated by $b \to c \bar c s$.  The current biggest discrepancy is in
$S_{\pi^0 K_S} = 0.33 \pm 0.21$, versus a nominal value of $0.678 \pm 0.026$.
This could be due, for instance, to exchange of a new $Z'$ masquerading as an
electroweak penguin.~\cite{GR07}  The study of $b \to s \ell^+ \ell^-$ and
searches at the Tevatron and LHC will see or bound $Z'$ effects.
Forward-backward asymmetries can be quite sensitive to $Z'$'s.~\cite{CDFZP,%
JLRZP}  One will be able to study such asymmetries at the LHC by passing to
non-zero pseudorapidity $\eta$.~\cite{Langacker:1984dc}

The $b \to s \ell^+ \ell^-$ decays show no anomalous behavior so far.%
~\cite{Scuri}  Belle/BaBar differ a bit and CDF agrees with BaBar with ${\cal
B}(B^0 \to K^{*0} \mu^+ \mu^-) = (0.82 \pm 0.31 \pm 0.10) \times 10^{-6}$, and
with Belle with ${\cal B}(B^+ \to K^+ \mu^+ \mu^-) = (0.60 \pm 0.15 \pm 0.04)
\times 10^{-6}$.

(3) It is encouraging to see the results searches for a right-handed
$W$:~\cite{CDFtop,CDFWR} $M_{W_R} > (790,760)$ MeV for $M_{W_R} (<,>)
M_{\nu_R}$.
The case of a right-handed $\nu_R$ heavier than $M_{W_R}$, in particular,
means that one must search for $W_R$ in the hadronic channel $t \bar b$.%
~\cite{JLRWR}

\section{Dark matter in many forms}

Ordinary matter exists in several stable forms:  $p$, $n$ (when incorporated
into nuclei), $e^-$, three flavors of neutrinos [$\tau(\nu_{2,3}) \gg \tu$].
We could expect dark matter (5--6 $\times$ ordinary matter) to exhibit at least
as much variety, for example if its quantum numbers are associated with a big
gauge group largely shielded from current observations.~\cite{Rosner:2005ec}
``Mirror particles,'' reviewed extensively by Okun,~\cite{Okun:2006eb} are
one example of this possibility.

There are at least two well-motivated dark matter candidates already (axions
and neutralinos). Axion dark matter has not received the attention it deserves.
RF cavity searches are going slowly; there is a large range of frequencies
still to be scanned with enough sensitivity.  Some variants of supersymmetry 
have long-lived next-to-lightest superpartners, decaying to the lightest
superpartners over a detectable distance.  Charged and neutral quasi-stable
candidates~\cite{Rizzi} could be split by so little that they charge-exchange
with the detector, implying new tracking signatures.

Dark matter could have non-zero charges purely in a hidden sector and thus be
invisible to all but gravitational probes.  Such opportunities might be
provided by the LISA detector.~\cite{Adams:2004pk}

Experience with hadron physics may help us deal with unexpected dark matter
forms and interactions.  This could be so, for example, if investigations at
the TeV scale uncover a new strongly-interacting sector, as expected in some
theories of dynamical electroweak symmetry breaking.

\section{Outlook}

Impressive measurements from BaBar, Belle, CDF, CERN NA48, CLEO, D0, KLOE,
RHIC, and other experiments have provided much fuel for theoretical
interpretations at this conference.  The understanding of hadron physics plays
a key role.  Much knowledge about fundamental electroweak interactions relies
on separating out the strong interactions.  Methods include theoretical
calculations (pQCD, SCET) and correlation of measurements through flavor
symmetry.  Conversely, low-energy hadron physics has benefitted greatly from
weak interactions; $K$, $D$, $B$ decays have provided information on $\pi \pi$
scattering, $\sigma$ and other scalar mesons, and patterns of final-state
interactions which go beyond what perturbative methods can anticipate.

Experiments at the Tevatron have shown that one can do excellent flavor physics
in a hadronic environment.  We look forward to fruitful results from LHCb on
$B_s \to \mu \mu$, CP violation in $B_s \to J/\psi \phi$, and many other
topics.

Higgs boson searches are gaining in both sensitivity and breadth; gaps are
being plugged.  In addition to the discovery of the Higgs at the LHC (unless
Fermilab finds it first!), we can look forward to measurements of $\sigma_T$,
flavor, top, Higgs, new particles and forces.

Discussions of a super-B-factory, possibly near Frascati, are maturing.%
~\cite{SuperB}  Such a
machine might solve the $b \to s$ penguin problem once and for all.  With a
luminosity approaching 100 times current values, it would permit tagging
with fully reconstructed $B$'s all those final states now studied with partial
tags.  Upgrades of KEK-B and LHCb also are being contemplated.  Finally,
neutrino studies \cite{Kayser} (near-term and more ambitious) and the ILC are
also on our horizon.  Our field has much to look forward to in
the coming decades.

\section*{Acknowledgments}
I would like to thank Jean and Kim Tr{\^a}n Thanh V{\^a}n, our gracious
hosts for these wonderful meetings; the Organizing Committee and Secretariat
for smooth arrangements; the hotel staff, for making us welcome, comfortable,
and well-fed; the funding agencies (EU, NSF) for supporting the attendance of
many participants; and all the speakers for their contributions to this
informative and enjoyable Moriond Workshop.  Sheldon Stone made a number of
useful comments on the manuscript.
This work was supported in part by the United States Department of Energy
through Grant No.\ DE FG02 90ER40560.


\section*{References}
\begin{thebibliography}{99}

\bibitem{Rolandi} G. Rolandi, this Conference.

\bibitem{D0top} L. Christofek [for the D0 Collaboration], this Conference.

\bibitem{CLEOfD} M.~Artuso {\it et al.} [CLEO Collaboration],
  Phys.\ Rev.\ Lett.\ {\bf 95}, 251801 (2005).

\bibitem{lattfD} C.~Aubin {\it et al.},
  Phys.\ Rev.\ Lett.\ {\bf 95}, 122002 (2005).

\bibitem{Artuso:2007zg} M.~Artuso {\it et al.} [CLEO Collaboration],
  arXiv:0704.0629 [hep-ex], submitted to PRL;
  T.~K.~Pedlar {\it et al.} [CLEO Collaboration],
  arXiv:0704.0437 [hep-ex], submitted to PR D.

\bibitem{Stone:ICHEP} S. Stone [CLEO Collaboration], arXiv:hep-ex/0610026,
presented at ICHEP 06 (Moscow, Russia, 2006), Proceedings edited by A. N.
Sissakian and G. A.  Kozlov, to be published by World Scientific, 2007.

\bibitem{Aubert:2006sd} B.~Aubert {\it et al.} [BABAR Collaboration],
  arXiv:hep-ex/0607094, submitted to PRL.

\bibitem{Abulencia:2006ze} A.~Abulencia {\it et al.} [CDF Collaboration],
  Phys.\ Rev.\ Lett.\ {\bf 97}, 242003 (2006).

\bibitem{Rosner:1990xx} J. L. Rosner, Phys.\ Rev.\ D {\bf 42}, 3732 (1990).

\bibitem{Okamotoxi} M.~Okamoto, PoS {\bf LAT2005}, 013 (2006),
arXiv:hep-lat/0510113.

\bibitem{D0mix} V. M. Abazov {\it et al.} [D0 Collaboration],
Phys.\ Rev.\ Lett.\ {\bf 97}, 021802 (2006).

\bibitem{D0life} V. M. Abazov {\it et al.} [D0 Collaboration],
arXiv:hep-ex/0702030; R. Jesik [for the D0 Collaboration], this Conference.

\bibitem{Lenz:2006hd} A.~Lenz and U.~Nierste,
  arXiv:hep-ph/0612167, based on NLO calculations in
  M.~Beneke, G.~Buchalla, C.~Greub, A.~Lenz and U.~Nierste,
  Phys.\ Lett.\ B {\bf 459}, 631 (1999), and
  M.~Beneke, G.~Buchalla, A.~Lenz and U.~Nierste,
  Phys.\ Lett.\ B {\bf 576}, 173 (2003).

\bibitem{Goloutvin} A. Goloutvin, this Conference.

\bibitem{John} M. John, this Conference.

\bibitem{Lin} E. Lin, this Conference.

\bibitem{Lu} C.-D. L\"u, this Conference.  See also A. Ali, {\it et al.},
arXiv:hep-ph/0703162.

\bibitem{Gronau:1994rj}
M.~Gronau, O.~F.~Hernandez, D.~London and J.~L.~Rosner,
Phys.\ Rev.\ D {\bf 50}, 4529  (1994); {\it ibid} {\bf 52}, 6374 (1995).

\bibitem{DH} N.~G.~Deshpande and X.~G.~He, Phys.\ Rev.\ Lett.\ {\bf 75}, 1703
(1995); X.~G.~He, Eur.\ Phys.\ J.\ C {\bf 9}, 443 (1999).

\bibitem{GR} M.~Gronau and J.~L.~Rosner, Phys.\ Rev.\ Lett.\ {\bf 76}, 1200
(1996); A.~S.~Dighe, M.~Gronau and J.~L.~Rosner, Phys.\ Rev.\ D {\bf 54},
3309 (1996).

\bibitem{rates} R.~Fleischer and T.~Mannel, Phys.\ Rev.\ D {\bf 57}, 2752
(1998); M.~Gronau and J.~L.~Rosner, Phys.\ Rev.\ D {\bf 57}, 6843 (1998);
A.~J.~Buras and R.~Fleischer, Eur.\ Phys.\ J.\ C {\bf 11}, 93 (1999); {\it
ibid.} {\bf 16}, 97 (2000); M.~Neubert and J.~L.~Rosner, Phys.\ Lett.\ B {\bf
441}, 403 (1998); Phys.\ Rev.\ Lett.\ {\bf 81}, 5076 (1998); M. Gronau, D.
Pirjol and T. M. Yan, Phys.\ Rev.\ D {\bf 60}, 034021 (1999); M.~Gronau and
J.~L.~Rosner, Phys.\ Rev.\ D {\bf 65}, 013004 (2002) [Erratum-ibid.\ D {\bf
65}, 079901 (2002)]; Phys.\ Lett.\ B {\bf 572}, 43 (2003).

\bibitem{fourCPsr} D.~Atwood and A.~Soni, Phys.\ Rev.\ D {\bf 58} 036005 (1998);
M.~Gronau, Phys.\ Lett.\ B {\bf 627}, 82 (2005); M.~Gronau and J.~L.~Rosner,
Phys.\ Rev.\ D {\bf 74}, 057503 (2006).

\bibitem{threeCPsr} M.~Gronau and J.~L.~Rosner,
  Phys.\ Rev.\  D {\bf 71}, 074019 (2005).

\bibitem{hfag} Heavy Flavor Averaging Group, hep-ex/0603003, as updated
periodically at \\ {\tt http://www.slac.stanford.edu/xorg/hfag/}.

\bibitem{ibreak} M.~Gronau, Y.~Grossman, G.~Raz and J.~L.~Rosner,
  Phys.\ Lett.\ B {\bf 635}, 207 (2006).

\bibitem{Aubert:2007mb} B.~Aubert {\it et al.} [BaBar Collaboration],
 arXiv:hep-ex/0702043, submitted to PRL.

\bibitem{pVpP} H. J. Lipkin, Phys.\ Rev.\ Lett.\ {\bf 46}, 1307 (1981); Phys.\
Lett.\ B {\bf 254}, 247 (1991); {\bf 415}, 186 (1997); {\bf 433}, 117 (1998);
M.~Gronau and J.~L.~Rosner, Phys.\ Rev.\ D {\bf 61}, 073008 (2000).

\bibitem{Dighe:1995pd} A.~S.~Dighe, I.~Dunietz, H.~J.~Lipkin and J.~L.~Rosner,
  Phys.\ Lett.\ B {\bf 369}, 144 (1996).

\bibitem{Scuri} F. Scuri, this Conference.

\bibitem{Medvedeva:2007af} T.~Medvedeva {\it et al.} [Belle Collaboration],
  arXiv:0704.2652 [hep-ex].

\bibitem{Wang:2007as} M.~Z.~Wang {\it et al.} [Belle Collaboration],
  arXiv:0704.2672 [hep-ex].

\bibitem{MZWang} M.-Z. Wang, this Conference.

\bibitem{JRexot} J.~L.~Rosner,
  Phys.\ Rev.\ D {\bf 69}, 094014 (2004)

\bibitem{GRsr} M. Gronau and J. L. Rosner, Phys.\ Lett.\ B {\bf 644}, 237
(2007).

\bibitem{BaBsinbeta} B. Aubert {\it et al.} [BaBar Collaboration],
arXiv:hep-ex/0703021, submitted to PRL.

\bibitem{Belsinbeta} K.~F.~Chen {\it et al.}  [Belle Collaboration],
  Phys.\ Rev.\ Lett.\ {\bf 98}, 031802 (2007).

\bibitem{Stocchi} A. Stocchi, this Conference.

\bibitem{BaDh} B. Aubert {\it et al.} [BABAR Collaboration],
  arXiv:hep-ex/0703019, submitted to PRL.

\bibitem{sin2bdiff} M. Gronau and J. L. Rosner, Phys\ Rev.\ D {\bf 74}, 093003
(2006).

\bibitem{MGJR07} M. Gronau and J. L. Rosner, 2007, in preparation.

\bibitem{MGJR04} M. Gronau and J. L. Rosner, Phys.\ Lett.\ B {\bf 595}, 339
(2004).

\bibitem{BaSC} B. Aubert {\it et al.} [BaBar Collaboration],
arXiv:hep-ex/0703016, submitted to PRL.

\bibitem{BeSC} K.~Abe {\it et al.} [Belle Collaboration],
  arXiv:hep-ex/0608035v2, submitted to PRL.

\bibitem{Fleischer:2007} R.~Fleischer, S.~Recksiegel and F.~Schwab,
  arXiv:hep-ph/0702275.

\bibitem{Becher} T. Becher, this Conference.

\bibitem{freq} F. James, this Conference; F. Le Diberder, this Conference.

\bibitem{Bay} G. Cowan, this Conference.

\bibitem{deedum} J. Charles {\it et al.} [CKMfitter Collaboration],
arXiv:hep-ph/0607246 and hep-ph/0703073v2;
M. Bona {\it et al.} [UTfit Collaboration], arXiv:hep-ph/0701207.

\bibitem{canc} A.~F.~Falk, Y.~Grossman, Z.~Ligeti and A.~A.~Petrov,
  Phys.\ Rev.\ D {\bf 65}, 054034 (2002);
  A.~F.~Falk, Y.~Grossman, Z.~Ligeti, Y.~Nir and A.~A.~Petrov,
  Phys.\ Rev.\  D {\bf 69}, 114021 (2004).

\bibitem{CLEOdKpi} W.~M.~Sun [for the CLEO Collaboration],
  AIP Conf.\ Proc.\ {\bf 842}, 693 (2006) [arXiv:hep-ex/0603031];
  D.~M.~Asner {\it et al.} [CLEO Collaboration],
  Int.\ J.\ Mod.\ Phys.\  A {\bf 21}, 5456 (2006)
  [arXiv:hep-ex/0607078].

\bibitem{nodKpi} R. L. Kingsley, S. B. Treiman, F. Wilczek and A. Zee,
Phys.\ Rev.\ D {\bf 11}, 1919 (1975); M. B. Voloshin, V. I. Zakharov and L. B.
Okun, Pis'ma v ZhETF {\bf 21}, 403 (1975) [Sov.\ Phys.--JETP Letters {\bf 21},
183 (1975); L. Wolfenstein, Phys.\ Rev.\ Lett.\ {\bf 75}, 2460 (1995);
M. Gronau, Y. Grossman, and J. L. Rosner, Phys.\ Lett.\ B {\bf 508}, 37 (2001).

\bibitem{Flood} K. Flood [for the BaBar Collaboration], this Conference.

\bibitem{Bamix} B.~Aubert {\it et al.} [BABAR Collaboration],
  arXiv:hep-ex/0703020, submitted to PRL.

\bibitem{Belife} M. Staric {\it et al.} [Belle Collaboration],
arXiv:hep-ex/0703036v2, presented at XLII Rencontres de Moriond, Electroweak
Interactions and Unified Theories, 10--17 March 2007.

\bibitem{Staric} M. Staric [for the Belle Collaboration], this Conference.

\bibitem{Bemix} K. Abe {\it et al.} [Belle Collaboration], arXiv:0704.1000
[hep-ex].

\bibitem{mixth} M. Ciuchini {\it et al.}, arXiv:hep-ph/0703204; Y. Nir,
arXiv:hep-ph/0703235; P. Ball, arXiv:hep-ph/0703245; V. Shevchenko, this
Conference.

\bibitem{NA48} B. Bloch-Devaux [for the NA48 Collaboration], this Conference.

\bibitem{Leut} H. Leutwyler, this Conference.

\bibitem{Blanc} F. Blanc [for the BaBar Collaboration], this Conference.

\bibitem{BL} G.~P.~Lepage and S.~J.~Brodsky,
  Phys.\ Rev.\ D {\bf 22}, 2157 (1980);
  S.~J.~Brodsky and G.~P.~Lepage,
  Phys.\ Rev.\ D {\bf 24}, 2848 (1981).

\bibitem{Ambrosino:2006gk} F.~Ambrosino {\it et al.} [KLOE Collaboration],
  arXiv:hep-ex/0612029, submitted to PL B.

\bibitem{DiMicco} B. Di Micco [for the KLOE Collaboration], this Conference.

\bibitem{JR83} J. L. Rosner, Phys.\ Rev.\ D {\bf 27}, 1101 (1983).

\bibitem{Wang} Z. Wang, this Conference.

\bibitem{Ma} H. Ma, this Conference.

\bibitem{Dubrovin} M. Dubrovin, this Conference.

\bibitem{Mitchell} R. Mitchell, this Conference.

\bibitem{Uehara} S. Uehara, this Conference.  See also S.~Uehara {\it et al.}
  [Belle Collaboration],
  Phys.\ Rev.\ Lett.\ {\bf 96}, 082003 (2006).

\bibitem{Aubert:2006je} B.~Aubert {\it et al.} [BABAR Collaboration],
  Phys.\ Rev.\ Lett.\ {\bf 97}, 232001 (2006).

\bibitem{Zhang} J. Zhang, this Conference.

\bibitem{Aubert:2007bt} B.~Aubert {\it et al.} [BABAR Collaboration],
  arXiv:hep-ex/0703030.

\bibitem{JROm} J. L. Rosner, Phys.\ Rev.\ D {\bf 52}, 6461 (1995).

\bibitem{chiral} M.~A.~Nowak, M.~Rho and I.~Zahed,
  Phys.\ Rev.\ D {\bf 48}, 4370 (1993)
  W. A. Bardeen and C. T. Hill, Phys.\ Rev.\ D {\bf 49}, 409 (1994).

\bibitem{Grenier} P. Grenier, this Conference.

\bibitem{Abe:2006xm} K.~Abe {\it et al.} [Belle Collaboration],
  arXiv:hep-ex/0608031.

\bibitem{Aubert:2006mh} B.~Aubert [BABAR Collaboration],
  Phys.\ Rev.\ Lett.\ {\bf 97}, 222001 (2006).

\bibitem{radex2857} E.~van Beveren and G.~Rupp, Phys.\ Rev.\ Lett.\ {\bf 97},
202001 (2006); F.~E.~Close, C.~E.~Thomas, O.~Lakhina and E.~S.~Swanson,
Phys.\ Lett.\ B {\bf 647}, 159 (2007).

\bibitem{3D2857} P.~Colangelo, F.~De Fazio and S.~Nicotri, Phys.\ Lett.\ B {\bf
642}, 48 (2006).

\bibitem{CLEO1} S.~Anderson {\it et al.} [CLEO Collaboration],
  Nucl.\ Phys.\ A {\bf 663}, 647 (2000).

\bibitem{Belle01} K.~Abe {\it et al.} [Belle Collaboration],
  Phys.\ Rev.\ D {\bf 69}, 112002 (2004).

\bibitem{FOCUS0} J.~M.~Link {\it et al.} [FOCUS Collaboration],
  Phys.\ Lett.\ B {\bf 586}, 11 (2004).

\bibitem{CDFsigb} CDF Collaboration, arXiv:hep-ex/0701056.

\bibitem{Filthaut} F. Filthaut, this Conference.

\bibitem{Acosta:2005mq} D. Acosta {\it et al.} [CDF Collaboration],
Phys.\ Rev.\ Lett. {\bf 96}, 202001 (2006).

\bibitem{Kuhn} J. H. K\"uhn, this Conference.

\bibitem{Rosner:2006yk} J. L. Rosner, Phys.\ Rev.\ D {\bf 75}, 013009 (2007).

\bibitem{Fano} U. Fano, Nuovo Cim.\ {\bf 12}, 154 (1935) [Translation: J. Res.\
Natl. Inst.\ Stand.\ Technol.\ {\bf 110}, 583 (2005)]; Phys.\ Rev.\ {\bf 124},
1866 (1961).

\bibitem{Wigner} E. P. Wigner, Phys.\ Rev.\ {\bf 73}, 1002 (1948).

\bibitem{Feshbach} H.~Feshbach, Ann. Phys.\ (N.Y.) {\bf 5}, 357 (1958).

\bibitem{Bedip} K. Abe {\it et al.} [Belle Collaboration],
arXiv:hep-ex/0608018.

\bibitem{Ba2S} B. Aubert {\it et al.} [BaBar Collaboration],
arXiv:hep-ex/0610057, submitted to PRL.

\bibitem{JRth} J. L. Rosner, Phys.\ Rev.\ D {\bf 74}, 076006 (2006).

\bibitem{Bath} B. Aubert {\it et al.} [BaBar Collaboration], Phys.\ Rev.\ D
{\bf 74}, 091103 (2006).

\bibitem{moments} J.~H.~Kuhn and M.~Steinhauser,
Nucl.\ Phys.\ B {\bf 619}, 588 (2001) [Erratum-ibid.\ B {\bf 640}, 415 (2002)].

\bibitem{CLEOLLb} D.~Besson {\it et al.} [CLEO Collaboration],
  Phys.\ Rev.\ D {\bf 71}, 012004 (2005).

\bibitem{PS} A. Pineda and A. Signer, PR D {\bf 73}, 111501 (2006).

\bibitem{Haisch} U. Haisch, this Conference.

\bibitem{Penin} A. Penin, this Conference.

\bibitem{HQprod} S. Miglioranzi, T. K\"uhl, and M. Campanelli, talks at this
Conference.

\bibitem{Kwong89} W. Kwong, L. H. Orr, and J. L. Rosner, Phys.\ Rev.\ D
{\bf 40}, 1453 (1989).

\bibitem{Andreev} V. Andreev, this Conference.

\bibitem{Krauss} F. Krauss, this Conference.

\bibitem{Richardson} P. Richardson, this Conference.

\bibitem{Kumano} S. Kumano, this Conference.  See also M. Hirai, S. Kumano,
T.-H. Nagai, and K. Sudoh, arXiv:hep-ph/0702250.

\bibitem{Varman} D. Varman, this Conference.

\bibitem{Badger} S. Badger, this Conference.

\bibitem{Zanderighi} G. Zanderighi, this Conference.

\bibitem{Cacciari} M. Cacciari, this Conference.

\bibitem{Terashi} K. Terashi, this Conference.

\bibitem{Cwiok} M. Cwiok, this Conference.

\bibitem{Norniella} O. Norniella, this Conference.

\bibitem{Salam} G. Salam, this Conference.

\bibitem{jetsHI} J. Putschke and C. Salgado, talks at this Conference.

\bibitem{WMG} E. Nurse, this Conference.

\bibitem{GW} J. L. Rosner, M.~P.~Worah and T.~Takeuchi, Phys.\ Rev.\ D
{\bf 49}, 1363 (1994).

\bibitem{top} U. Ba{\ss}ler, M. Wang, and U. Husemann, talks at this
Conference; Tevatron Electroweak Working Group, arXiv:hep-ex/0703034.

\bibitem{PW} M. Peskin and J. Wells, Phys.\ Rev.\ D {\bf 64}, 093003 (2001).

\bibitem{JRHiggs} J. L. Rosner, Phys.\ Rev.\ D {\bf 65}, 073026 (2002).

\bibitem{CDFtop} W. Wagner [for the CDF Collaboration], this Conference.

\bibitem{Wurthwein} F. W\"urthwein, this Conference.

\bibitem{Higgs} G. Landsberg, this Conference.

\bibitem{Kilminster} B. Kilminster, this Conference.

\bibitem{Meissner} K. Meissner, this Conference.

\bibitem{Allwood} S. Allwood, this Conference.

\bibitem{SL} E.~Fullana and M.~A.~Sanchis-Lozano,
  arXiv:hep-ph/0702190, and earlier references therein.

\bibitem{Kniehl:2003ap}
  B.~A.~Kniehl, A.~A.~Penin, A.~Pineda, V.~A.~Smirnov and M.~Steinhauser,
  Phys.\ Rev.\ Lett.\  {\bf 92}, 242001 (2004).

\bibitem{Pineda} A. Pineda, this Conference.

\bibitem{Weiglein} G. Weiglein, this Conference.

\bibitem{compass} G. Brona and A. Vrona, talks at this Conference.

\bibitem{star} K. Kowalik, this Conference.

\bibitem{HERAPV} S. U. Noor, this Conference.

\bibitem{HERAsf} T. Danielson, this Conference.

\bibitem{HERApp} T. Hreus, this Conference.

\bibitem{JCS} J. Casalderrey-Solana, this Conference.

\bibitem{Deines-Jones:1996fe}P.~Deines-Jones {\it et al.} [KLMM Collaboration],
  Phys.\ Rev.\ C {\bf 53}, 3044 (1996).

\bibitem{Cunqueiro} L. Cunqueiro, this Conference.

\bibitem{Belghobsi} Z. Belghobsi, this Conference.

\bibitem{HBT} P. Romatschke and P. Chaloupka, talks at this Conference.
See also P. Romatschke, arXiv:nucl-th/0701032.

\bibitem{charmint} A. Dion, this Conference.

\bibitem{partsat} G. Stephens and D. D'Enterria, talks at this Conference.

\bibitem{qsupp} A. Rakotozafindrabe, this Conference.

\bibitem{Jegerlehner} F. Jegerlehner, arXiv:hep-ph/0703125.

\bibitem{GR07} M.~Gronau and J.~L.~Rosner,
  arXiv:hep-ph/0702193.

\bibitem{CDFZP} A. Abulencia {\it et al.} [CDF Collaboration], Phys.\ Rev.\ D
{\bf 96}, 211801 (2006).

\bibitem{JLRZP} J. L. Rosner, Phys.\ Rev.\  D {\bf 54}, 1078 (1996).

\bibitem{Langacker:1984dc} P.~Langacker, R.~W.~Robinett and J.~L.~Rosner,
  Phys.\ Rev.\ D {\bf 30}, 1470 (1984).

\bibitem{CDFWR} V.~M.~Abazov {\it et al.} [D0 Collaboration],
  Phys.\ Lett.\  B {\bf 641}, 423 (2006);
  A.~Abulencia {\it et al.} [CDF Collaboration],
  arXiv:hep-ex/0611022; S. Muanza, this Conference.

\bibitem{JLRWR} J. L. Rosner, Phys.\ Rev.\ D {\bf 42}, 241 (1990).

\bibitem{Rosner:2005ec} J.~L.~Rosner, arXiv:astro-ph/0509196.

\bibitem{Okun:2006eb} L.~B.~Okun,
  arXiv:hep-ph/0606202.

\bibitem{Rizzi} A. Rizzi, this Conference.

\bibitem{Adams:2004pk} A.~W.~Adams and J.~S.~Bloom, arXiv:astro-ph/0405266.

\bibitem{SuperB} See {\tt http://www.pi.infn.it/SuperB} for information.

\bibitem{Kayser} B. Kayser, this Conference.

\end{thebibliography}
\end{document}